\documentclass[showpacs,showkeys,amsmath,amssymb,preprint]{revtex4}
\usepackage{epsfig,dcolumn,bm}

\begin{document}

\title{$NK$ and $\Delta K$ states in the chiral
SU(3) quark model}

\author{F. Huang}
\affiliation{ CCAST (World Laboratory), P.O. Box 8730, Beijing
100080, PR China \\ Institute of High Energy Physics, P.O. Box
918-4, Beijing 100049, PR China\footnote{Mailing address.} \\
Graduate School of the Chinese Academy of Sciences, Beijing, PR
China}
\author{Z.Y. Zhang}
\affiliation{Institute of High Energy Physics, P.O. Box 918-4,
Beijing 100049, PR China}

\begin{abstract}
The isospin $I=0$ and $I=1$ kaon-nucleon $S$, $P$, $D$, $F$ wave
phase shifts are studied in the chiral SU(3) quark model by
solving the resonating group method (RGM) equation. The calculated
phase shifts for different partial waves are in agreement with the
experimental data. Furthermore, the structures of the $\Delta K$
states with $L=0$, $I=1$ and $I=2$ are investigated. We find that
the interaction between $\Delta$ and $K$ in the case of $L=0$,
$I=1$ is attractive, which is not like the situation of the $NK$
system, where the $S$-wave interactions between $N$ and $K$ for
both $I=0$ and $I=1$ are repulsive. Our numerical results also
show that when the model parameters are taken to be the same as in
our previous $NN$ and $YN$ scattering calculations, the $\Delta K$
state with $L=0$ and $I=1$ is a weakly bound state with about 2
MeV binding energy, while the one with $I=2$ is unbound in the
present one-channel calculation.
\end{abstract}



\preprint{nucl-th/0409029}

\maketitle

\section{Introduction}

In the framework of the constituent quark model, to understand the
source of the constituent quark mass, the spontaneous vacuum
breaking has to be considered, and as a consequence the coupling
between quark field and goldstone boson is introduced to restore
the chiral symmetry. In this sense, the chiral quark model can be
regarded as a quite reasonable and useful model to describe the
medium-range nonperturbative QCD effect. By generalizing the SU(2)
linear $\sigma$ model, a chiral SU(3) quark model is developed to
describe the system with strangeness \cite{zyz97}. This model has
been quite successful in reproducing the energies of the baryon
ground states, the binding energy of deuteron, the nucleon-nucleon
($NN$) scattering phase shifts of different partial waves, and the
hyperon-nucleon ($YN$) cross sections by performing the resonating
group method (RGM) calculations \cite{zyz97, lrd03}. Recently this
model has been extended to the systems with antiquarks to study
the baryon-meson interactions and meson-meson interactions. With
the antiquark ($\bar q$) in the meson brought in, the complexity
of the annihilation part in the interactions will appear. We
started our study from the $KN$ elastic scattering processes
because in the $KN$ system the annihilations to gluons and vacuum
are forbidden and the $u(d)\bar s$ can only annihilate to kaon
mesons. We calculated the $KN$ phase shifts of S and P partial
waves in the chiral SU(3) quark model and fortunately, we got
quite reasonable agreement with the experimental data when the
mixing between scalar mesons $\sigma_0$ and $\sigma_8$ is
considered \cite{fhu04}. All of these achievements encourage us to
extend our study to some higher partial waves of $KN$ scattering
and to some other $5q$ systems with an antiquark $\bar s$, such as
$\Delta K$, $NK^*$, and so on.

Actually, the $KN$ scattering process had aroused particular
interest in the past \cite{rbu90,nbl02,dha02,sle03,hjw03} and many
works have been devoted to this issue based on constituent quark
model. But up to now, most of them can not accurately reproduce
the $KN$ phase shifts up to $L=3$ in a sufficient consistent way.
In this work we perform a RGM calculation of $S$, $P$, $D$, $F$
wave $KN$ phase shifts in our chiral quark model. Comparing with
other's previous work \cite{sle03}, a RGM calculation including
$\sigma$ and $\pi$ boson exchanges, we obtain correct signs of
$S_{01}$, $P_{11}$, $P_{03}$, $D_{13}$, $D_{05}$, $F_{15}$,
$F_{07}$ waves, and for $P_{01}$, $D_{03}$, $D_{15}$ channels we
also get a considerable improvement on the theoretical phase
shifts in the magnitude.

About the structures of $\Delta K$ states, Sarkar {\it et al.}
have made a study on a baryon level in their recent work
\cite{ssa04}, and they obtained a $\Delta K$ resonance state with
$L=0$ and $I=1$ near the threshold. This state has also been
investigated by Kolomeitsev and Luta in the $\chi$-BS(3) approach
\cite{eek04}. In the present work we analyze the property of the
one-gluon-exchange (OGE) interaction in $\Delta K$ states with
$L=0$ and our results show that the OGE interaction is attractive
for $I=1$ while repulsive for $I=2$. When the model parameters are
taken to be the same as in our previous work which successfully
reproduced the $NN$ phase shifts and the $YN$ cross sections
\cite{zyz97,lrd03}, we find that the $\Delta K$ with $L=0$ and
$I=1$ is a weakly bound state with the binding energy about 2 MeV
in our present one-channel calculation.

The paper is organized as follows. In the next section the
framework of the chiral SU(3) quark model is briefly introduced.
The results of $KN$ phase shifts and $\Delta K$ states are shown
in Sec. III, where some discussions are made as well.
Finally, conclusions are given in Sec. IV.

\section{Formulation}

\subsection{The model}

As is well known, the nonperturbative QCD effect is very important
in light quark system.  To consider the low-momentum medium-range
nonperturbative QCD effect, a SU(2) linear $\sigma$ model
\cite{mge60,ffe93} is supposed to study the $NN$ interactions. In
order to extend the study to the systems with strangeness, we
generalized the idea of the SU(2) $\sigma$ model to the flavor
SU(3) case, in which a unified coupling between quarks and all
scalar and pseudoscalar chiral fields is introduced and the
constituent quark mass can be understood in principle as the
consequence of spontaneous chiral symmetry breaking of the QCD
vacuum \cite{zyz97}. With this generalization, the interacting
Hamiltonian between quarks and chiral fields can be written as
\begin{eqnarray}\label{hami}
H^{ch}_I = g_{ch} F({\bm q}^{2}) \bar{\psi} \left( \sum^{8}_{a=0}
\sigma_a \lambda_a + i \sum^{8}_{a=0} \pi_a \lambda_a \gamma_5
\right) \psi,
\end{eqnarray}
where $g_{ch}$ is the coupling constant between quark and
chiral-field, and $\lambda_{0}$ a unitary matrix. $\lambda_{1},
..., \lambda_{8}$ are the Gell-Mann matrix of the flavor SU(3)
group, $\sigma_{0},...,\sigma_{8}$ the scalar nonet fields and
$\pi_{0},..,\pi_{8}$ the pseudoscalar nonet fields. $F({\bm
q}^{2})$ is a form factor inserted to describe the chiral-field
structure \cite{ito90,amk91} and as usual, it is taken to be
\begin{eqnarray}\label{faca}
F({\bm q}^{2}) = \left(\frac{\Lambda^2}{\Lambda^2+{\bm
q}^2}\right)^{1/2},
\end{eqnarray}
with $\Lambda$ being the cutoff mass of the chiral field. Clearly,
${H}^{ch}_{I}$ is invariant under the infinitesimal chiral SU(3)
transformation.

From ${H}^{ch}_{I}$, the chiral-field-induced effective
quark-quark potentials can be derived, and their expressions are
given in the following:
\begin{eqnarray}
V_{\sigma_a}({\bm r}_{ij})=-C(g_{ch},m_{\sigma_a},\Lambda)
X_1(m_{\sigma_a},\Lambda,r_{ij}) [\lambda_a(i)\lambda_a(j)] +
V_{\sigma_a}^{\bm {l \cdot s}}({\bm r}_{ij}),
\end{eqnarray}
\begin{eqnarray}
V_{\pi_a}({\bm r}_{ij})=C(g_{ch},m_{\pi_a},\Lambda)
\frac{m^2_{\pi_a}}{12m_{q_i}m_{q_j}} X_2(m_{\pi_a},\Lambda,r_{ij})
({\bm \sigma}_i\cdot{\bm \sigma}_j) [\lambda_a(i)\lambda_a(j)]
+V_{\pi_a}^{ten}({\bm r}_{ij}),
\end{eqnarray}
and
\begin{eqnarray}
V_{\sigma_a}^{\bm {l \cdot s}}({\bm r}_{ij})&=&
-C(g_{ch},m_{\sigma_a},\Lambda)\frac{m^2_{\sigma_a}}{4m_{q_i}m_{q_j}}
\left\{G(m_{\sigma_a}r_{ij})-\left(\frac{\Lambda}{m_{\sigma_a}}\right)^3
G(\Lambda r_{ij})\right\} \nonumber\\
&&\times[{\bm L \cdot ({\bm \sigma}_i+{\bm
\sigma}_j)}][\lambda_a(i)\lambda_a(j)],
\end{eqnarray}
\begin{eqnarray}
V_{\pi_a}^{ten}({\bm r}_{ij})&=&
C(g_{ch},m_{\pi_a},\Lambda)\frac{m^2_{\pi_a}}{12m_{q_i}m_{q_j}}
\left\{H(m_{\pi_a}r_{ij})-\left(\frac{\Lambda}{m_{\pi_a}}\right)^3
H(\Lambda r_{ij})\right\} \nonumber\\
&&\times\left[3({\bm \sigma}_i \cdot \hat{r}_{ij})({\bm \sigma}_j
\cdot \hat{r}_{ij})-{\bm \sigma}_i \cdot {\bm
\sigma}_j\right][\lambda_a(i)\lambda_a(j)],
\end{eqnarray}
with
\begin{eqnarray}
C(g_{ch},m,\Lambda)=\frac{g^2_{ch}}{4\pi}
\frac{\Lambda^2}{\Lambda^2-m^2} m,
\end{eqnarray}
\begin{eqnarray}
\label{x1mlr} X_1(m,\Lambda,r)=Y(mr)-\frac{\Lambda}{m} Y(\Lambda
r),
\end{eqnarray}
\begin{eqnarray}
X_2(m,\Lambda,r)=Y(mr)-\left(\frac{\Lambda}{m}\right)^3 Y(\Lambda
r),
\end{eqnarray}
\begin{eqnarray}
Y(x)=\frac{1}{x}e^{-x},
\end{eqnarray}
\begin{eqnarray}
G(x)=\frac{1}{x}\left(1+\frac{1}{x}\right)Y(x),
\end{eqnarray}
\begin{eqnarray}
H(x)=\left(1+\frac{3}{x}+\frac{3}{x^2}\right)Y(x),
\end{eqnarray}
and $m_{\sigma_a}$ being the mass of the scalar meson and
$m_{\pi_a}$ the mass of the pseudoscalar meson.

In the chiral SU(3) quark model, the interaction induced by the
coupling of chiral field describes the nonperturbative QCD effect
of the low-momentum medium-distance range. To study the hadron
structure and hadron-hadron dynamics, one still needs to include
an effective one-gluon-exchange interaction $V^{OGE}_{ij}$ which
governs the short-range behavior,
\begin{eqnarray}
V^{OGE}_{ij}=\frac{1}{4}g_{i}g_{j}\left(\lambda^c_i\cdot\lambda^c_j\right)
\left\{\frac{1}{r_{ij}}-\frac{\pi}{2} \delta({\bm r}_{ij})
\left(\frac{1}{m^2_{q_i}}+\frac{1}{m^2_{q_j}}+\frac{4}{3}\frac{1}{m_{q_i}m_{q_j}}
({\bm \sigma}_i \cdot {\bm \sigma}_j)\right)\right\}+V_{OGE}^{\bm
l \cdot \bm s},
\end{eqnarray}
with
\begin{eqnarray}
V_{OGE}^{\bm l \cdot \bm
s}=-\frac{1}{16}g_ig_j\left(\lambda^c_i\cdot\lambda^c_j\right)
\frac{3}{m_{q_i}m_{q_j}}\frac{1}{r^3_{ij}}{\bm L \cdot ({\bm
\sigma}_i+{\bm \sigma}_j)},
\end{eqnarray}
and a confinement potential $V^{conf}_{ij}$ to provide the
nonperturbative QCD effect in the long distance,
\begin{eqnarray}
V_{ij}^{conf}=-a_{ij}^{c}(\lambda_{i}^{c}\cdot\lambda_{j}^{c})r_{ij}^2
-a_{ij}^{c0}(\lambda_{i}^{c}\cdot\lambda_{j}^{c}).
\end{eqnarray}

For the systems with an antiquark ${\bar s}$, the total
Hamiltonian can be written as \cite{fhu04}
\begin{eqnarray}
\label{hami5q}
H=\sum_{i=1}^{5}T_{i}-T_{G}+\sum_{i<j=1}^{4}V_{ij}+\sum_{i=1}^{4}V_{i\bar
5},
\end{eqnarray}
where $T_G$ is the kinetic energy operator of the center of mass
motion, and $V_{ij}$ and $V_{i\bar 5}$ represent the interactions
between quark-quark ($qq$) and quark-antiquark ($q{\bar q}$)
respectively,
\begin{eqnarray}
V_{ij}= V^{OGE}_{ij} + V^{conf}_{ij} + V^{ch}_{ij},
\end{eqnarray}
\begin{eqnarray}
V^{ch}_{ij}=\sum^{8}_{a=0}V_{\sigma_a}(\bm
r_{ij})+\sum^{8}_{a=0}V_{\pi_a} (\bm r_{ij}).
\end{eqnarray}
$V_{i \bar 5}$ in Eq. (\ref{hami5q}) includes two parts
\cite{fhu03,fhu04}: direct interactions and annihilation parts,
\begin{eqnarray}
V_{i\bar 5}=V^{dir}_{i\bar 5}+V^{ann}_{i\bar 5},
\end{eqnarray}
with
\begin{eqnarray}
V_{i\bar 5}^{dir}=V_{i\bar 5}^{conf}+V_{i\bar 5}^{OGE}+V_{i\bar
5}^{ch},
\end{eqnarray}
where
\begin{eqnarray}
V_{i\bar
5}^{conf}=-a_{i5}^{c}\left(-\lambda_{i}^{c}\cdot{\lambda_{5}^{c}}^*\right)r_{i\bar
5}^2
-a_{i5}^{c0}\left(-\lambda_{i}^{c}\cdot{\lambda_{5}^{c}}^*\right),
\end{eqnarray}
\begin{eqnarray}
V^{OGE}_{i\bar
5}&=&\frac{1}{4}g_{i}g_{s}\left(-\lambda^c_i\cdot{\lambda^c_5}^*\right)
\left\{\frac{1}{r_{i\bar 5}}-\frac{\pi}{2} \delta({\bm r}_{i\bar
5})
\left(\frac{1}{m^2_{q_i}}+\frac{1}{m^2_{s}}+\frac{4}{3}\frac{1}{m_{q_i}m_{s}}
({\bm \sigma}_i \cdot {\bm \sigma}_5)\right)\right\}  \nonumber \\
&&-\frac{1}{16}g_ig_s\left(-\lambda^c_i\cdot{\lambda^c_5}^*\right)
\frac{3}{m_{q_i}m_{q_5}}\frac{1}{r^3_{i\bar 5}}{\bm L \cdot ({\bm
\sigma}_i+{\bm \sigma}_5)},
\end{eqnarray}
and
\begin{eqnarray}
V_{i\bar{5}}^{ch}=\sum_{j}(-1)^{G_j}V_{i5}^{ch,j}.
\end{eqnarray}
Here $(-1)^{G_j}$ represents the G parity of the $j$-th meson. For
the $NK$ and $\Delta$K systems, $u(d)\bar{s}$ can only annihilate
into a kaon meson, i.e.,
\begin{eqnarray}
V_{i\bar 5}^{ann}=V_{ann}^{K},
\end{eqnarray}
with
\begin{eqnarray}
V_{ann}^{K}=C^K\left(\frac{1-{\bm \sigma}_q \cdot {\bm
\sigma}_{\bar{q}}}{2}\right)_{s}\left(\frac{2 + 3\lambda_q \cdot
\lambda^*_{\bar{q}}}{6}\right)_{c} \left(\frac{38+3\lambda_q \cdot
\lambda^*_{\bar q}}{18}\right)_{f}\delta({\bm r}),
\end{eqnarray}
where $C^K$ is treated as a parameter and we adjust it to fit the
mass of kaon meson.

\subsection{Determination of parameters}

We have three initial input parameters: the harmonic-oscillator
width parameter $b_u$, the up (down) quark mass $m_{u(d)}$, and
the strange quark mass $m_s$. These three parameters are taken to
be the usual values: $b_u=0.5$ fm, $m_{u(d)}=313$ MeV, and
$m_s=470$ MeV. By some special constraints, the other model
parameters are fixed in the following way. The chiral coupling
constant $g_{ch}$ is fixed by
\begin{eqnarray}
\frac{g^{2}_{ch}}{4\pi} = \left( \frac{3}{5} \right)^{2}
\frac{g^{2}_{NN\pi}}{4\pi} \frac{m^{2}_{u}}{M^{2}_{N}},
\end{eqnarray}
with $g^{2}_{NN\pi}/4\pi=13.67$ taken as the experimental value.
The masses of the mesons are also adopted to the experimental
values, except for the $\sigma$ meson, where its mass is treated
as an adjustable parameter. In our previous work it is taken to be
595 MeV \cite{lrd03} for $NN$ and $YN$ cases while 675 MeV for the
present $KN$ case. The cutoff radius $\Lambda^{-1}$ is taken to be
the value close to the chiral symmetry breaking scale
\cite{ito90,amk91,abu91,emh91}. After the parameters of chiral
fields are fixed, the one gluon exchange coupling constants
$g_{u}$ and $g_{s}$ can be determined by the mass splits between
$N$, $\Delta$ and $\Sigma$, $\Lambda$, respectively. The
confinement strengths $a^{c}_{uu}$, $a^{c}_{us}$, and $a^{c}_{ss}$
are fixed by the stability conditions of $N$, $\Lambda$, and
$\Xi$, and the zero point energies $a^{c0}_{uu}$, $a^{c0}_{us}$,
and $a^{c0}_{ss}$ by fitting the masses of $N$, $\Sigma$ and
$\overline{\Xi+\Omega}$, respectively.

{\small
\begin{table}[htb]
\caption{\label{para} Model parameters. The meson masses and the
cutoff masses: $m_{\sigma'}=980$ MeV, $m_{\kappa}=980$ MeV,
$m_{\epsilon}=980$ MeV, $m_{\pi}=138$ MeV, $m_K=495$ MeV,
$m_{\eta}=549$ MeV, $m_{\eta'}=957$ MeV, $\Lambda=1100$ MeV.}
\begin{center}
\begin{tabular*}{160mm}{@{\extracolsep\fill}cccc}
\hline\hline
  & For $NN$, $YN$ cases & \multicolumn{2}{c}{For $KN$ case}  \\ \cline{3-4}
  &  & $\theta^S=35.264^\circ$ & $\theta^S=-18^\circ$ \\
\hline
 $b_u$ (fm)  & 0.5 & 0.5 & 0.5 \\
 $m_u$ (MeV) & 313 & 313 & 313 \\
 $m_s$ (MeV) & 470 & 470 & 470 \\
 $g_u$     & 0.886 & 0.886 & 0.886 \\
 $g_s$     & 0.917 & 0.917 & 0.917 \\
 $m_\sigma$ (MeV) & 595 & 675 & 675 \\
 $a^c_{uu}$ (MeV/fm$^2$) & 48.1 & 52.4 & 55.2 \\
 $a^c_{us}$ (MeV/fm$^2$) & 60.7 & 72.3 & 68.4 \\
 $a^{c0}_{uu}$ (MeV)  & $-$43.6 & $-$50.4 & $-$55.1 \\
 $a^{c0}_{us}$ (MeV)  & $-$38.2 & $-$54.2 & $-$48.7 \\
\hline\hline
\end{tabular*}
\end{center}
\end{table}}

In the calculation, $\eta$ and $\eta'$ mesons are mixed by
$\eta_1$ and $\eta_8$ with the mixing angle $\theta^{PS}$ taken to
be the usual value $-23^\circ$. For the $KN$ case, we also
consider the mixing between $\sigma_0$ and $\sigma_8$. The mixing
angle $\theta^{S}$ is still an open issue because the structure of
$\sigma$ meson is unclear and controversial. We adopt two possible
values by which we can get reasonable $KN$ phase shifts. One is
$35.264^\circ$ which means that $\sigma$ and $\epsilon$ are
ideally mixed by $\sigma_0$ and $\sigma_8$, and the other is
$-18^\circ$ which is provided by Dai and Wu \cite{ybd03} based on
their recent investigation.

The three sets of model parameters are tabulated in Table
\ref{para}. The first column is for the case fitted by $NN$ and
$YN$ scattering, and the second and third columns are for the case
fitted by $KN$ scattering. For each set of parameters the octet
and decuplet baryons' masses can be well reproduced in our model
\cite{pns99,fhu04}.

\section{Results and discussions}

\subsection{$KN$ phase shifts}

\begin{figure}[htb]
\epsfig{file=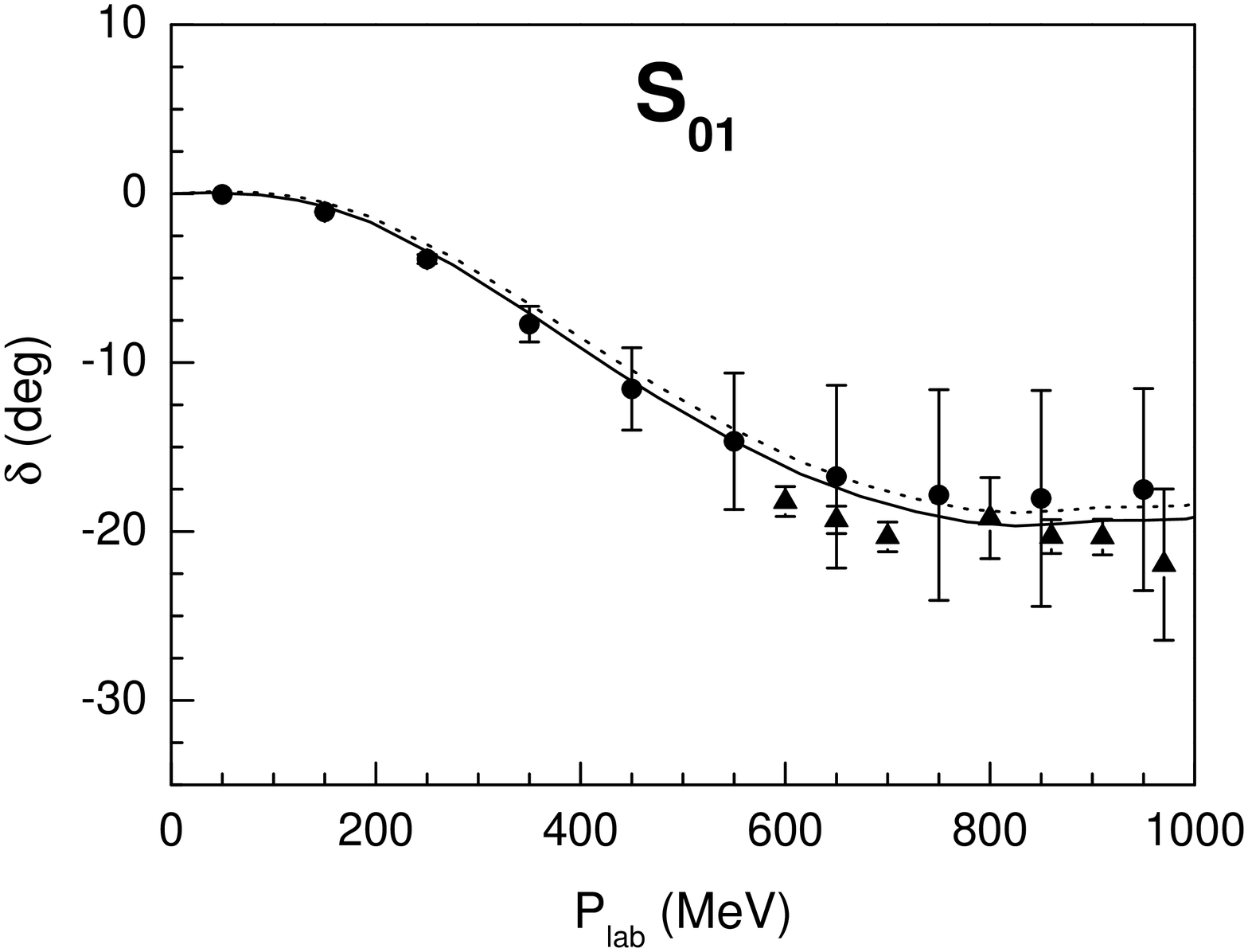,width=7.7cm}
\epsfig{file=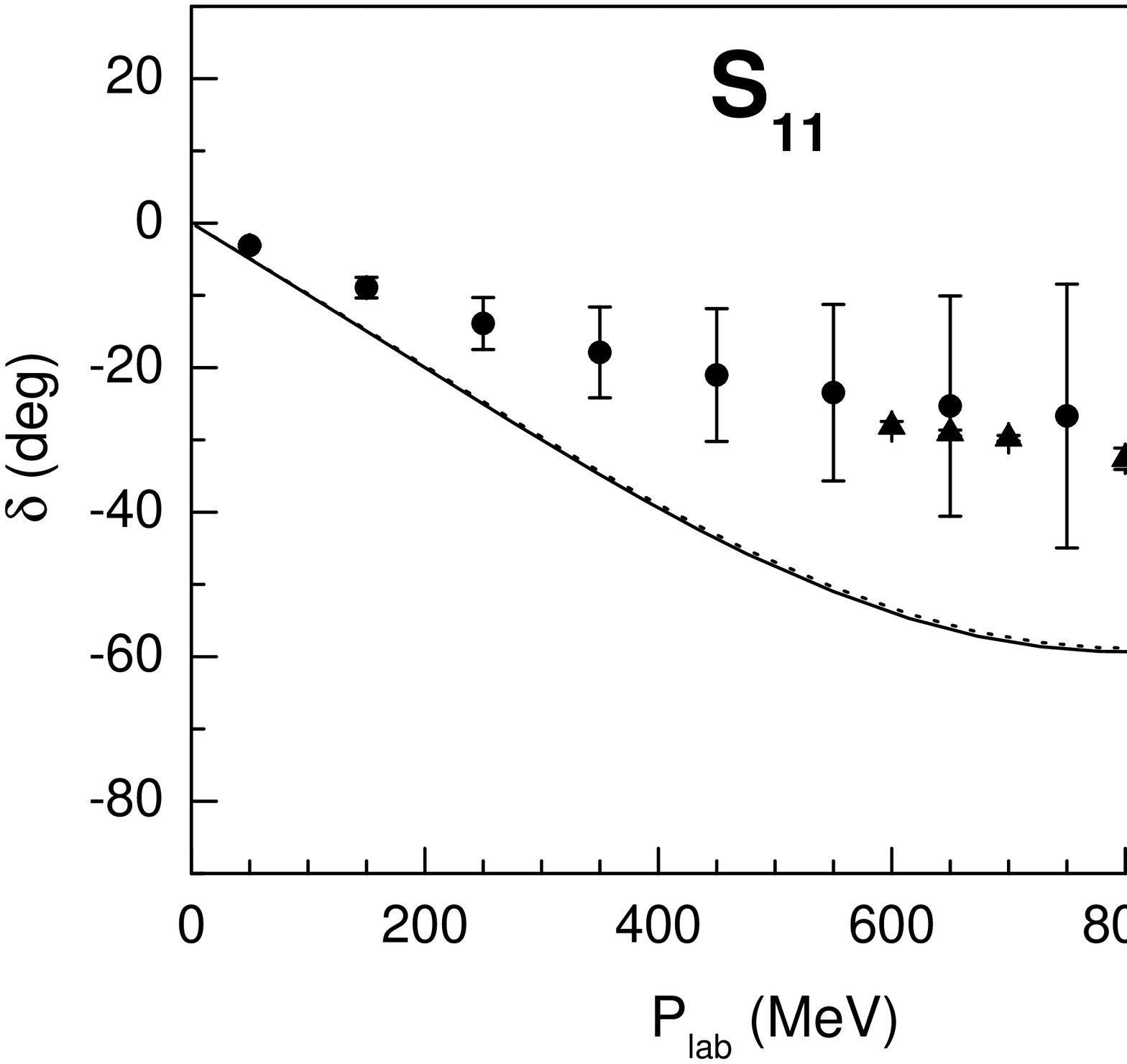,width=7.7cm} \vglue -2.3cm \caption{\small
\label{s0s1} $KN$ $S$-wave phase shifts as a function of the
laboratory momentum of kaon meson. The hole circles and the
triangles correspond respectively to the phase shifts analysis of
Hyslop {\it et al.} \cite{jsh92} and Hashimoto \cite{kha84}.}
\end{figure}

\begin{figure}[htb]
\epsfig{file=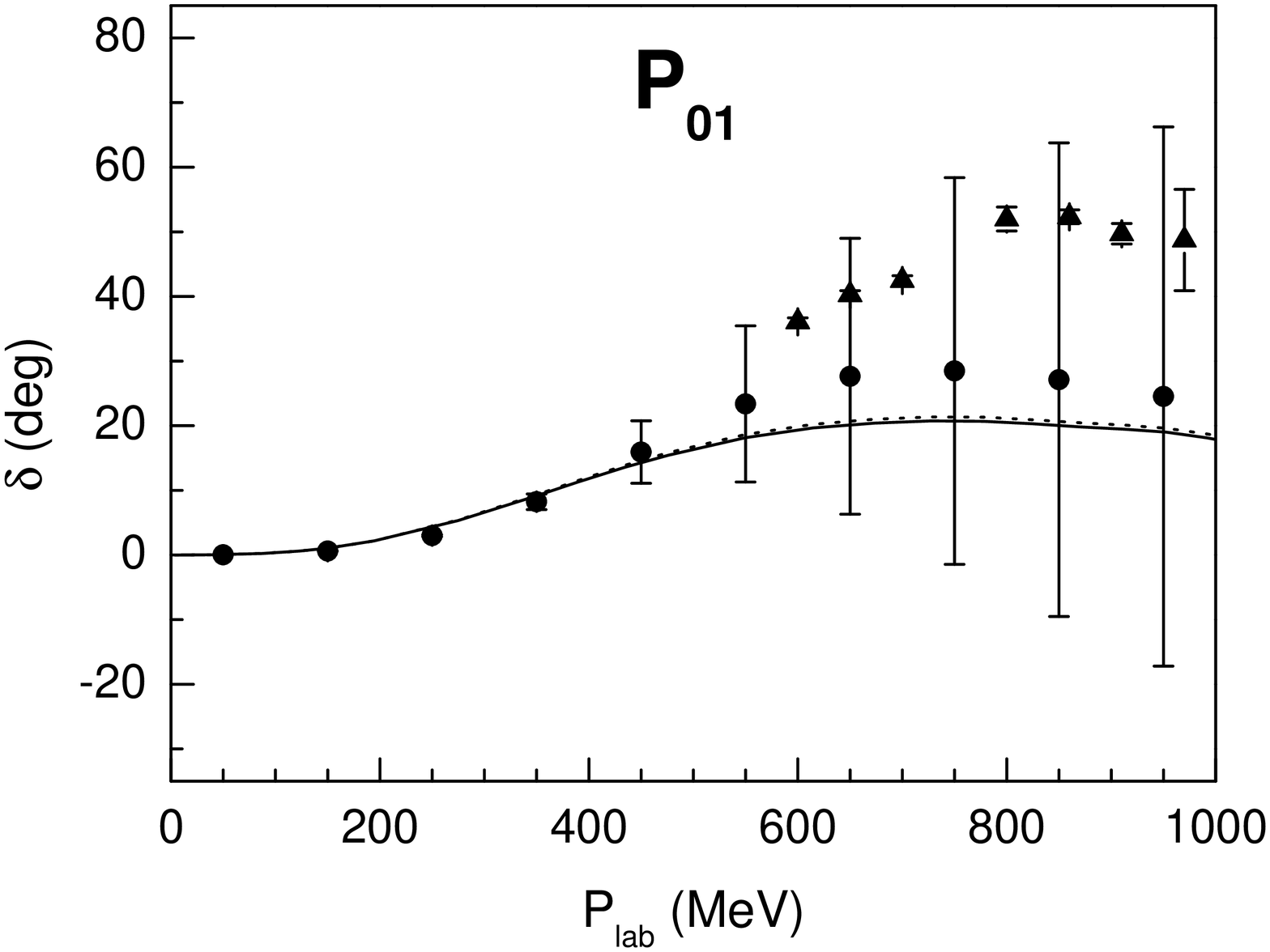,width=7.7cm}
\epsfig{file=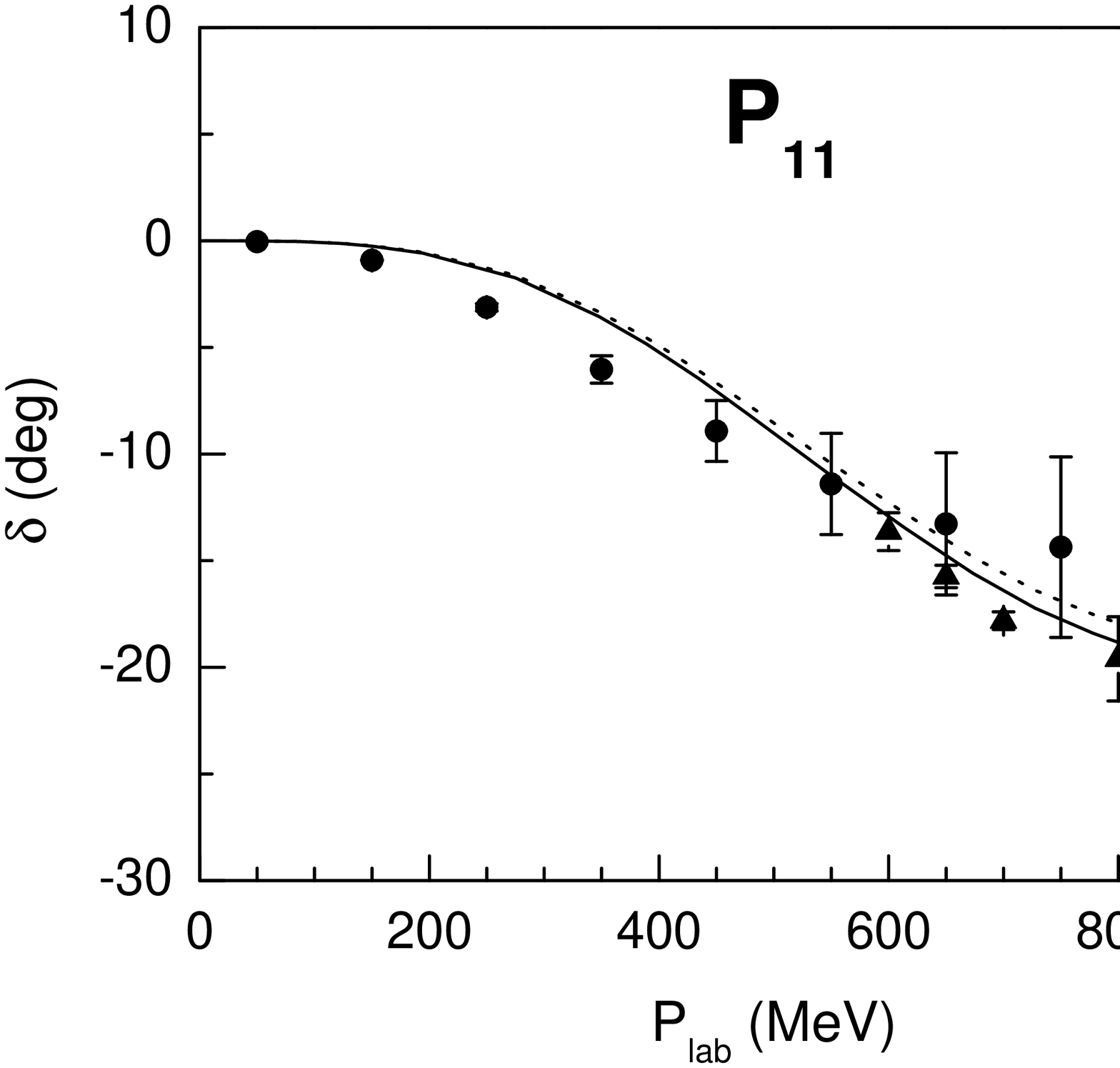,width=7.7cm}
\epsfig{file=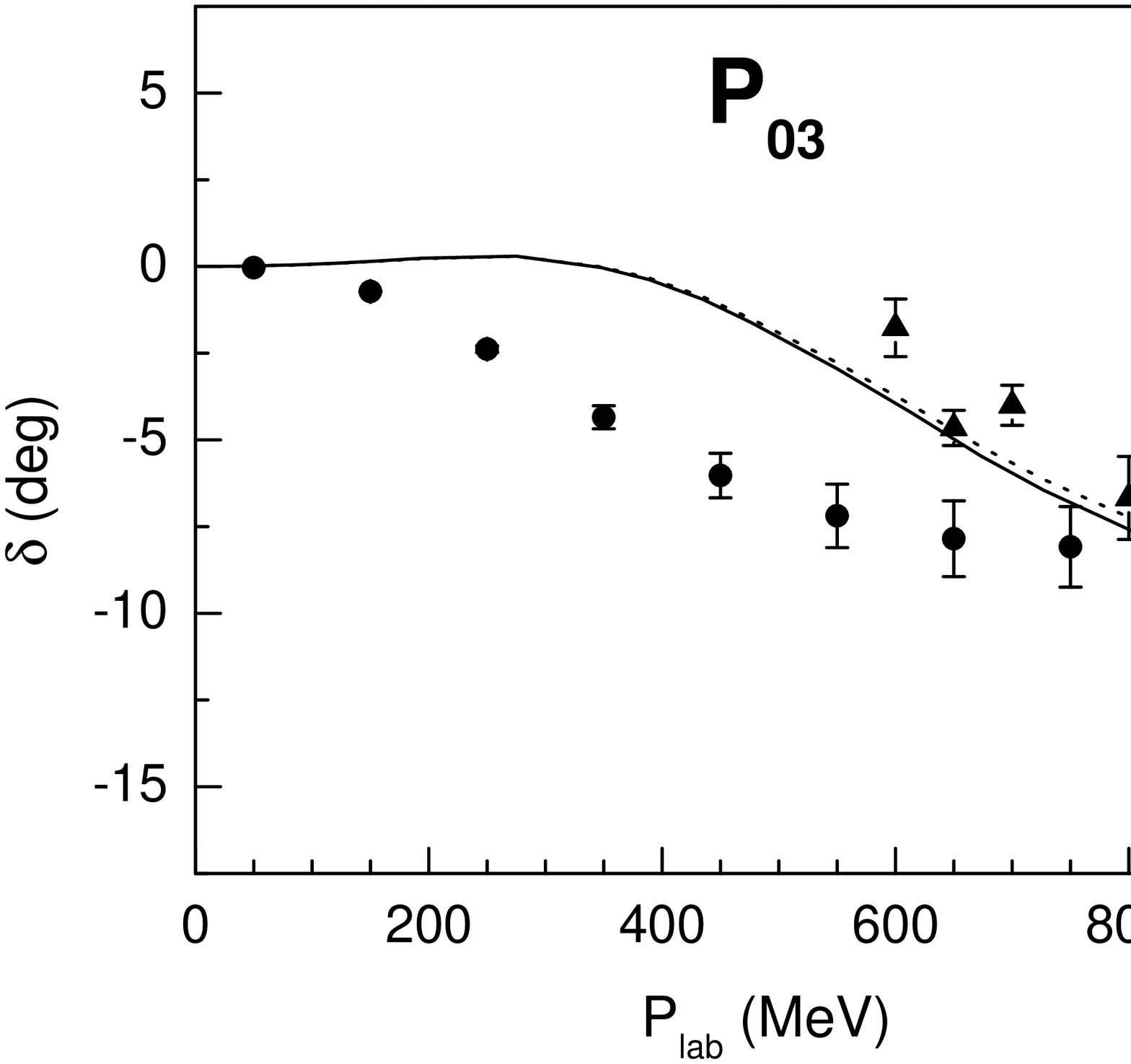,width=7.7cm}
\epsfig{file=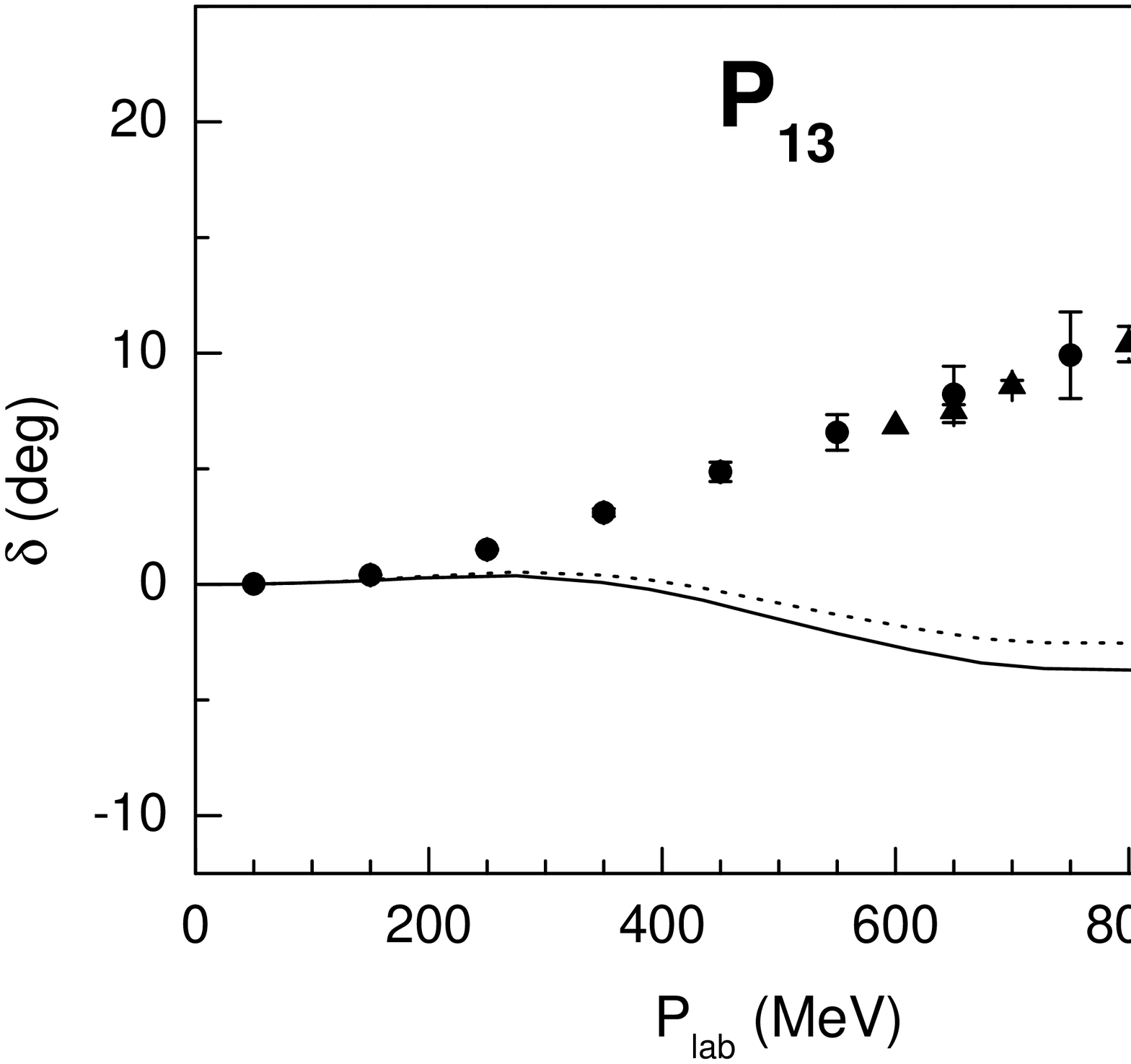,width=7.7cm} \vglue -2.3cm \caption{\small
\label{p0p1} $KN$ $P$-wave phase shifts. Same notation as in Fig.
\ref{s0s1}.}
\end{figure}

\begin{figure}[htb]
\epsfig{file=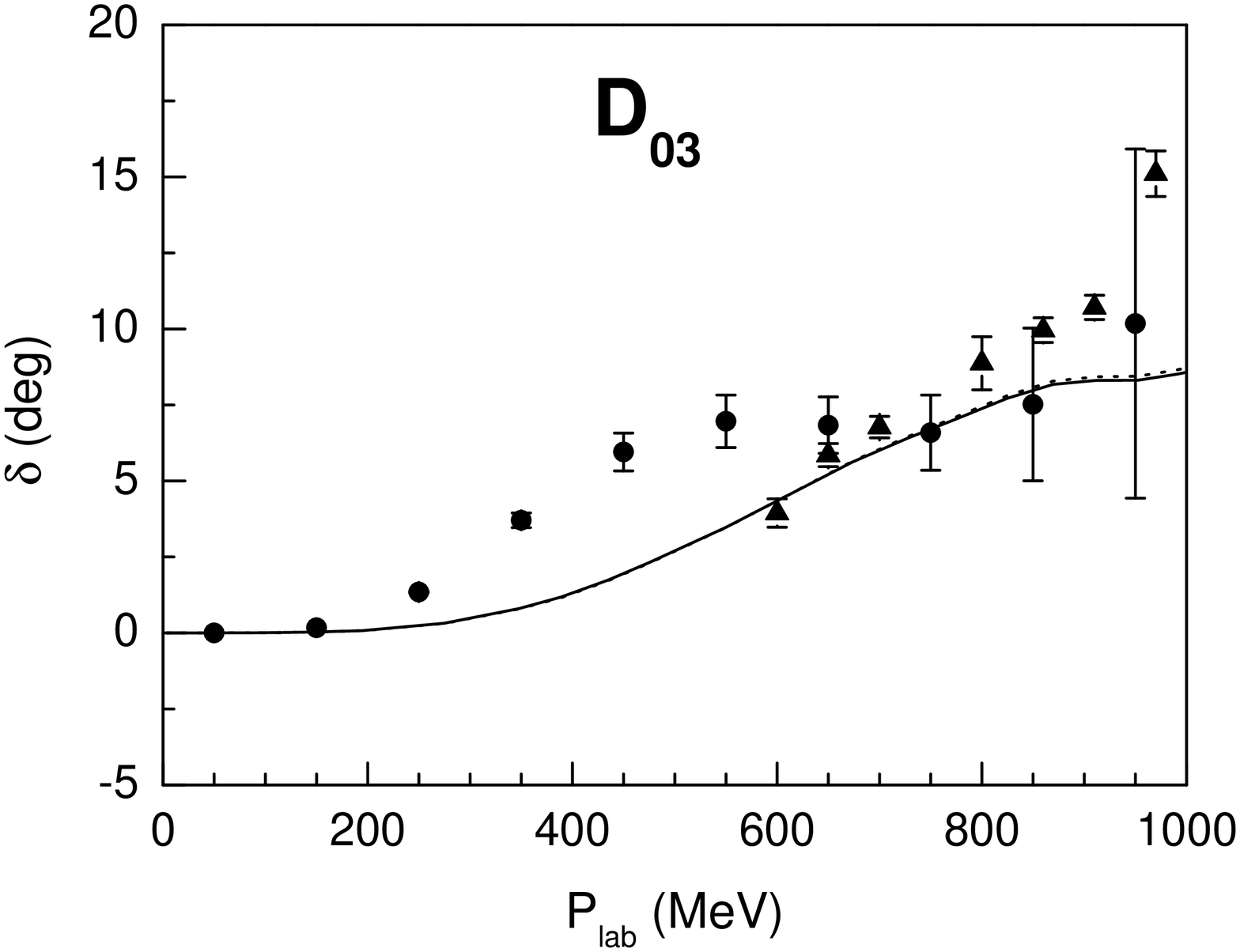,width=7.7cm}
\epsfig{file=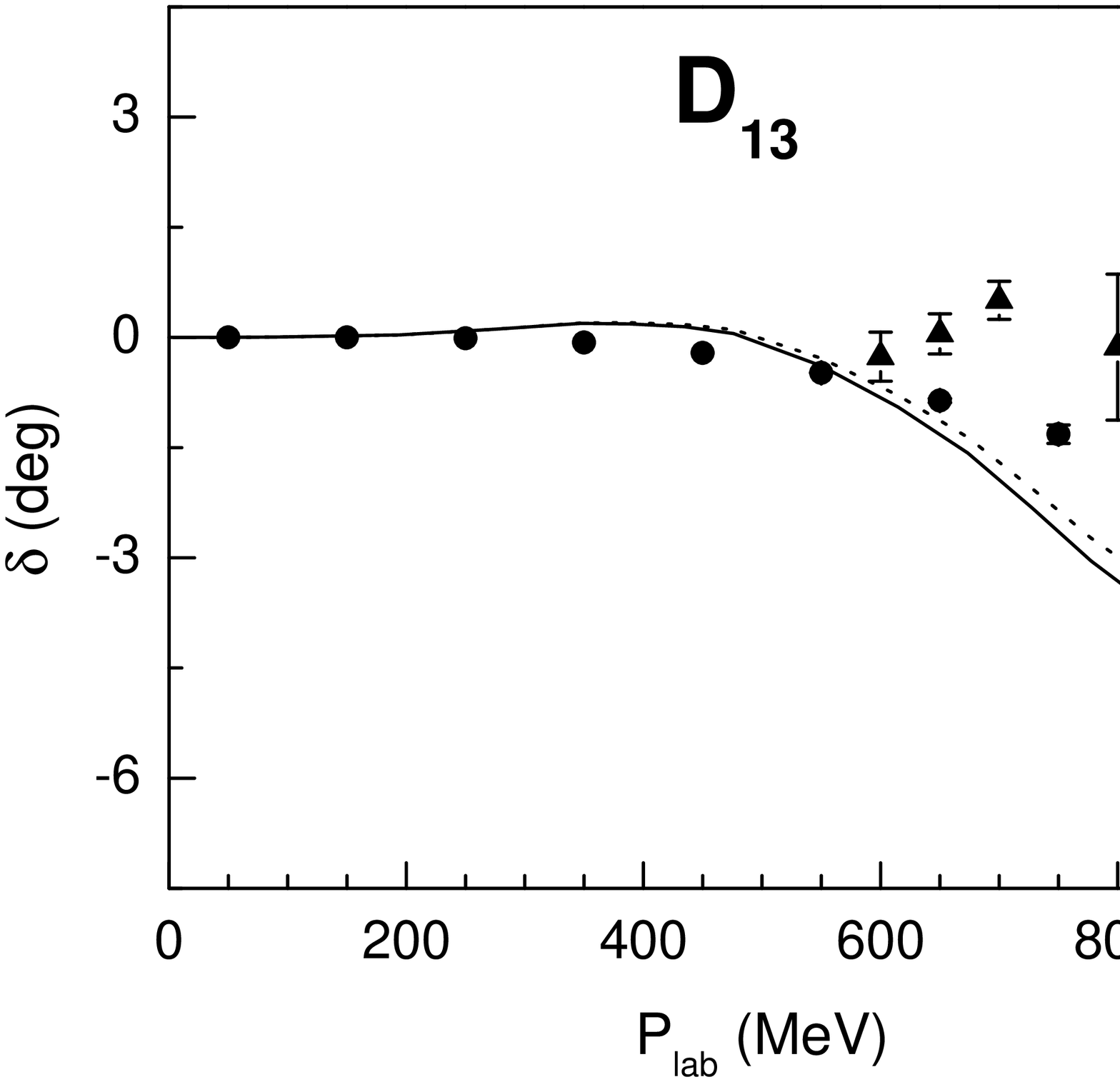,width=7.7cm}
\epsfig{file=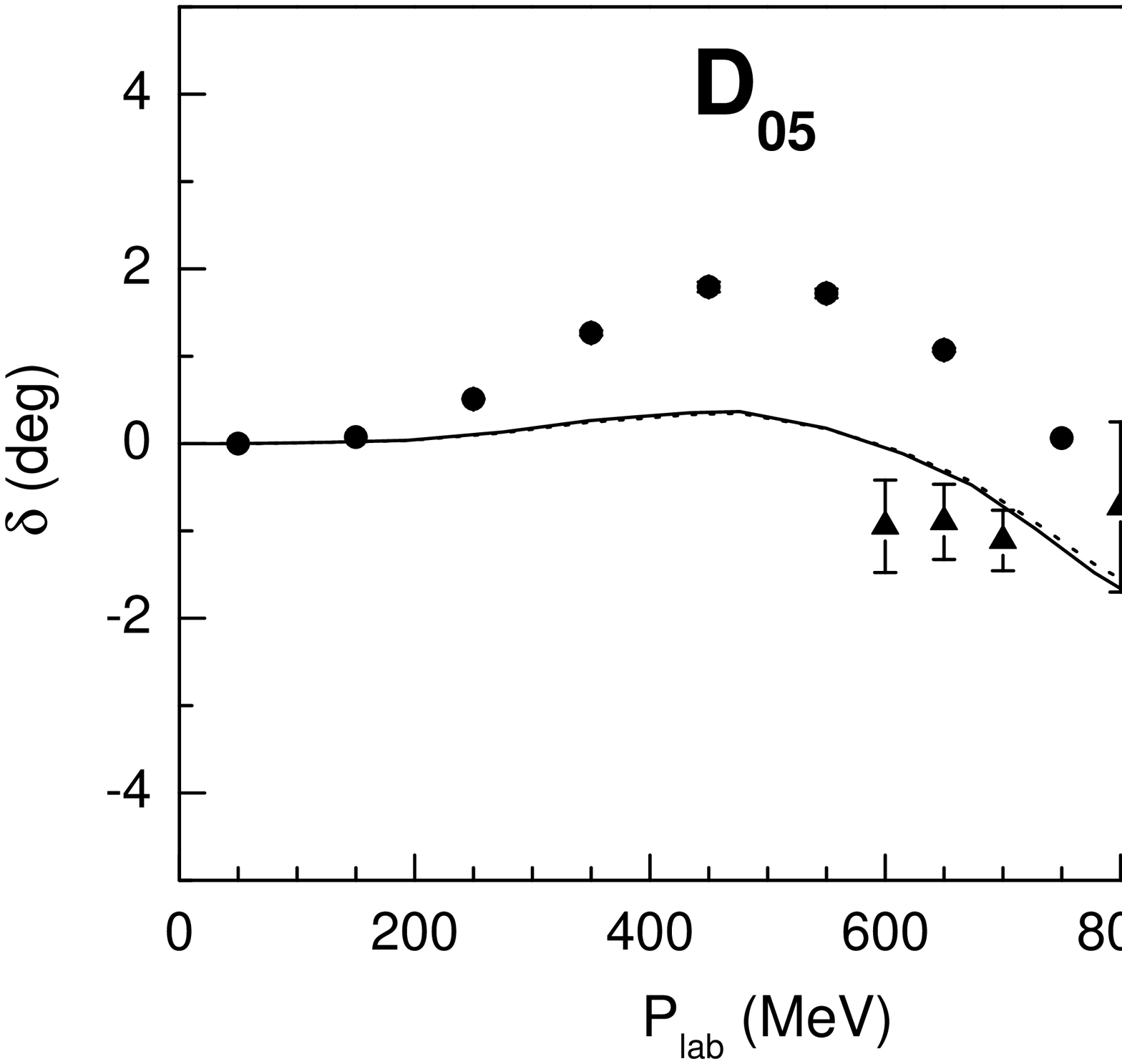,width=7.7cm}
\epsfig{file=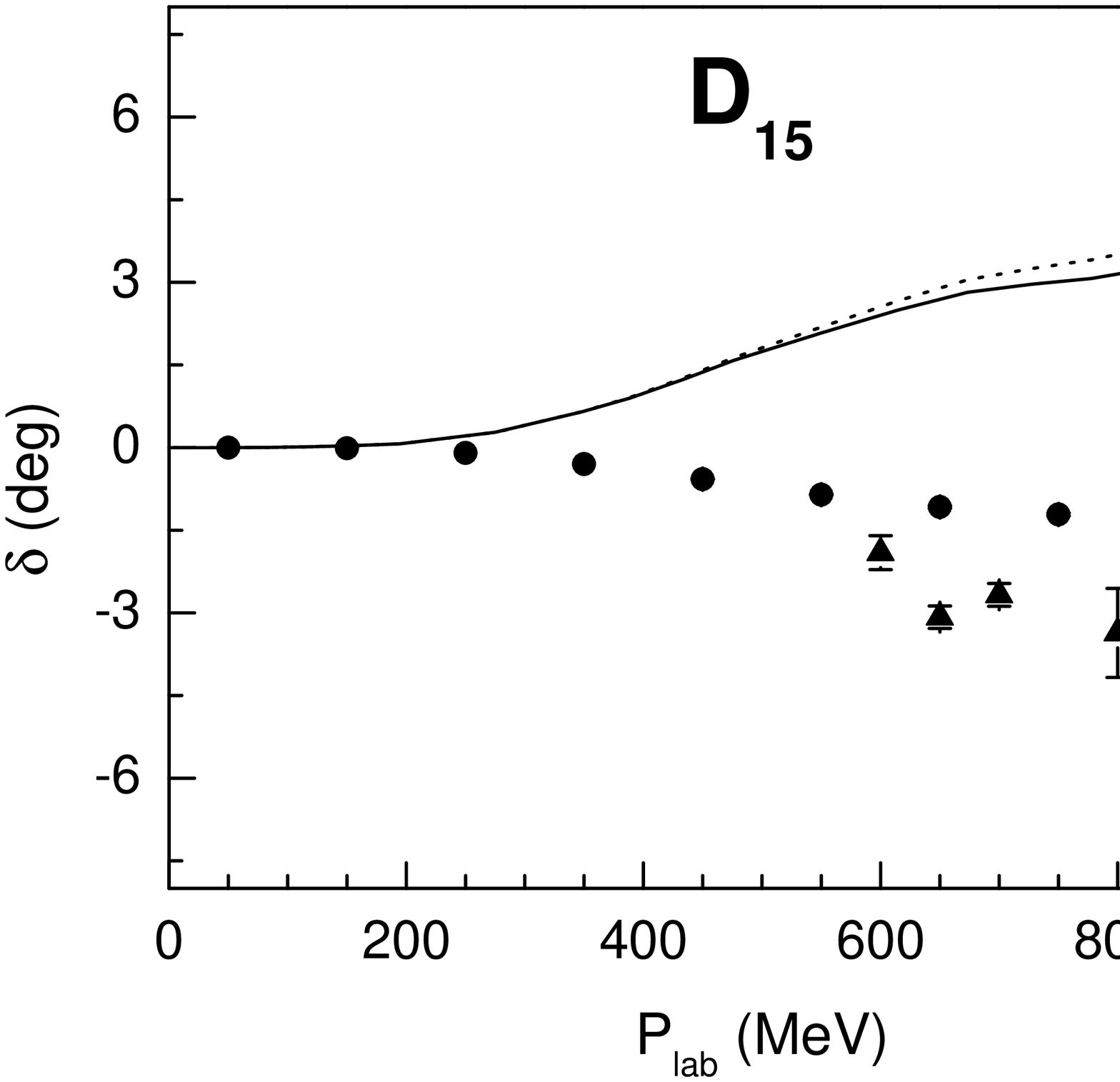,width=7.7cm} \vglue -2.3cm \caption{\small
\label{d0d1} $KN$ $D$-wave phase shifts. Same notation as in Fig.
\ref{s0s1}.}
\end{figure}

\begin{figure}[htb]
\epsfig{file=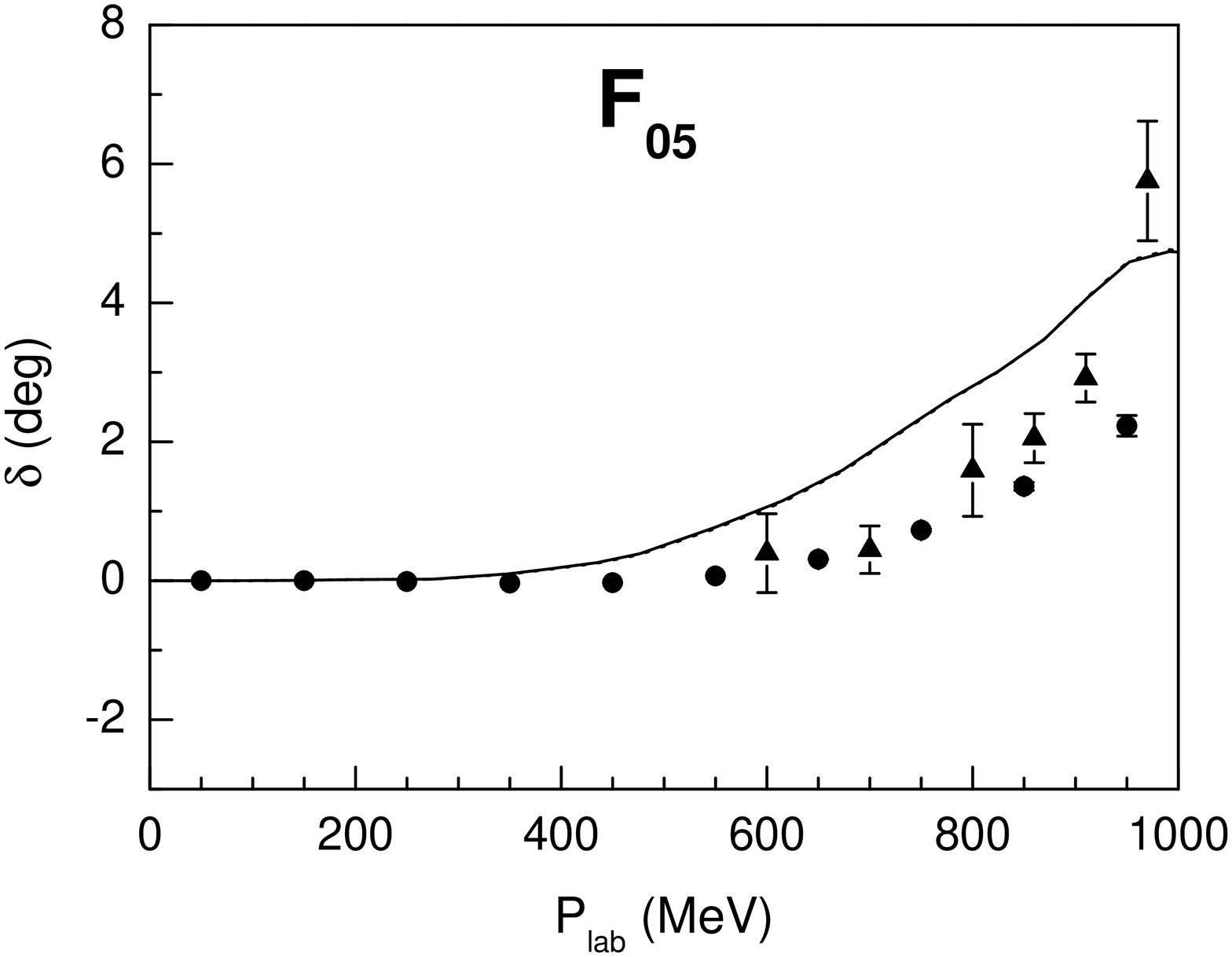,width=7.7cm}
\epsfig{file=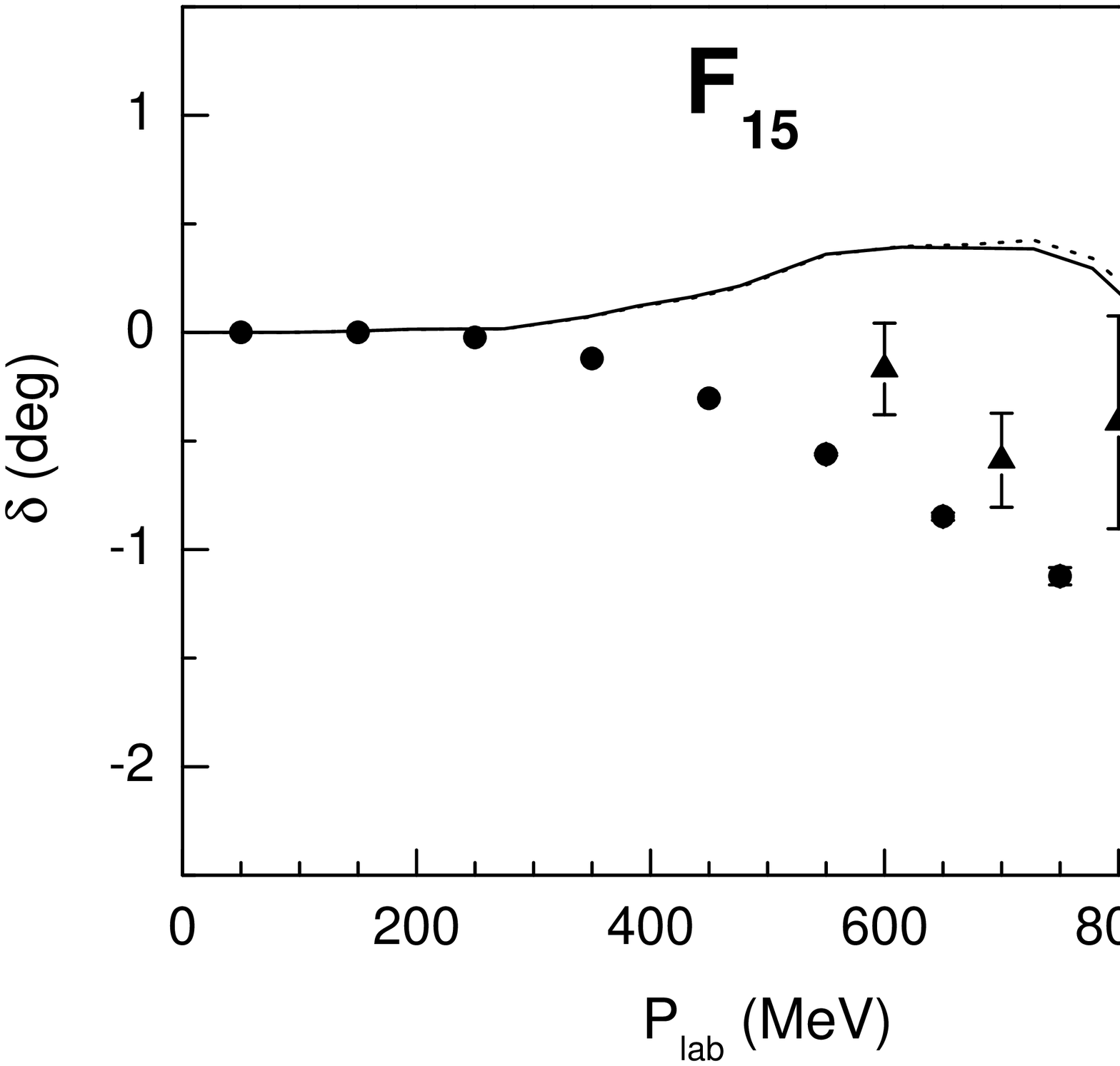,width=7.7cm}
\epsfig{file=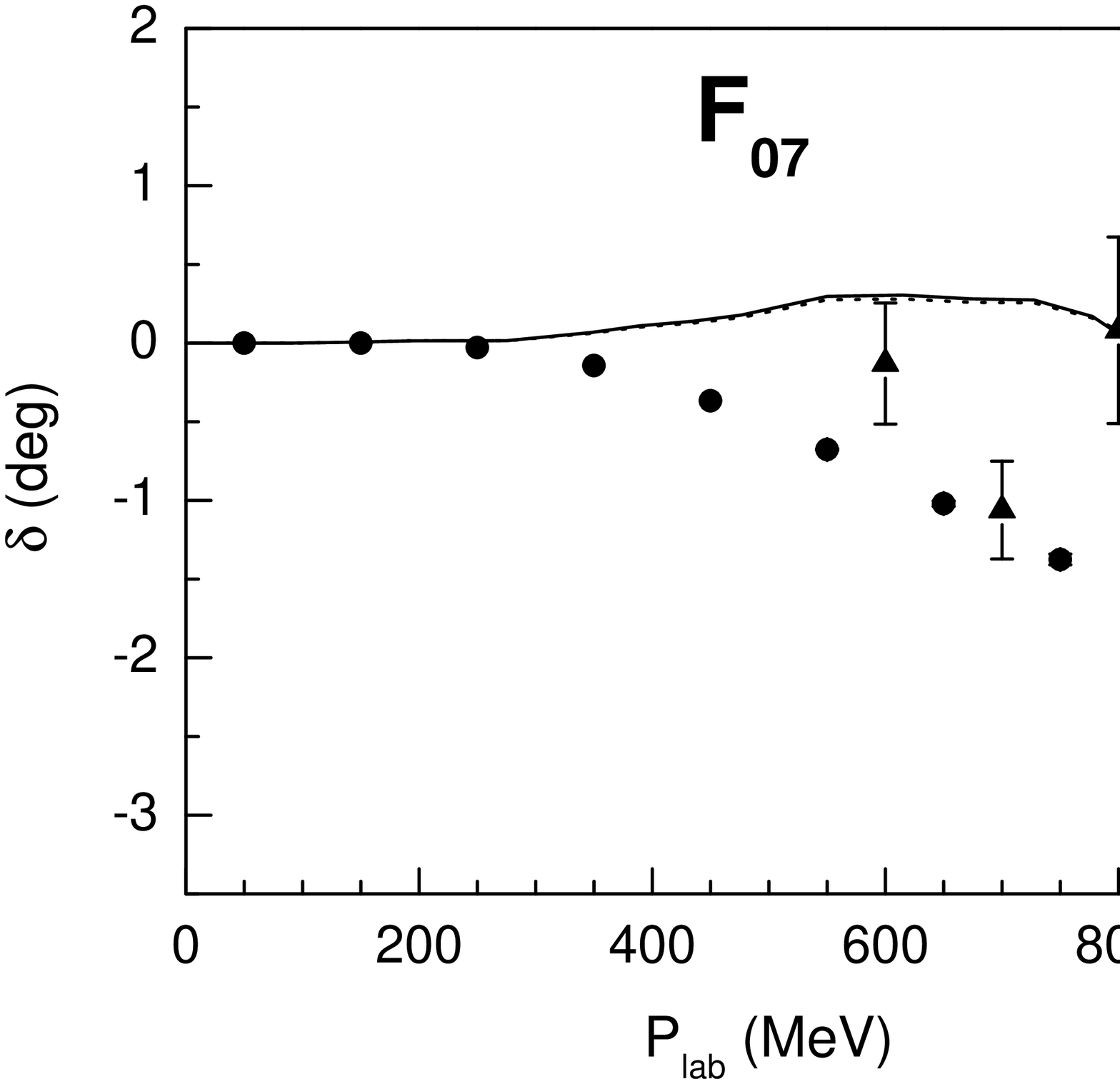,width=7.7cm}
\epsfig{file=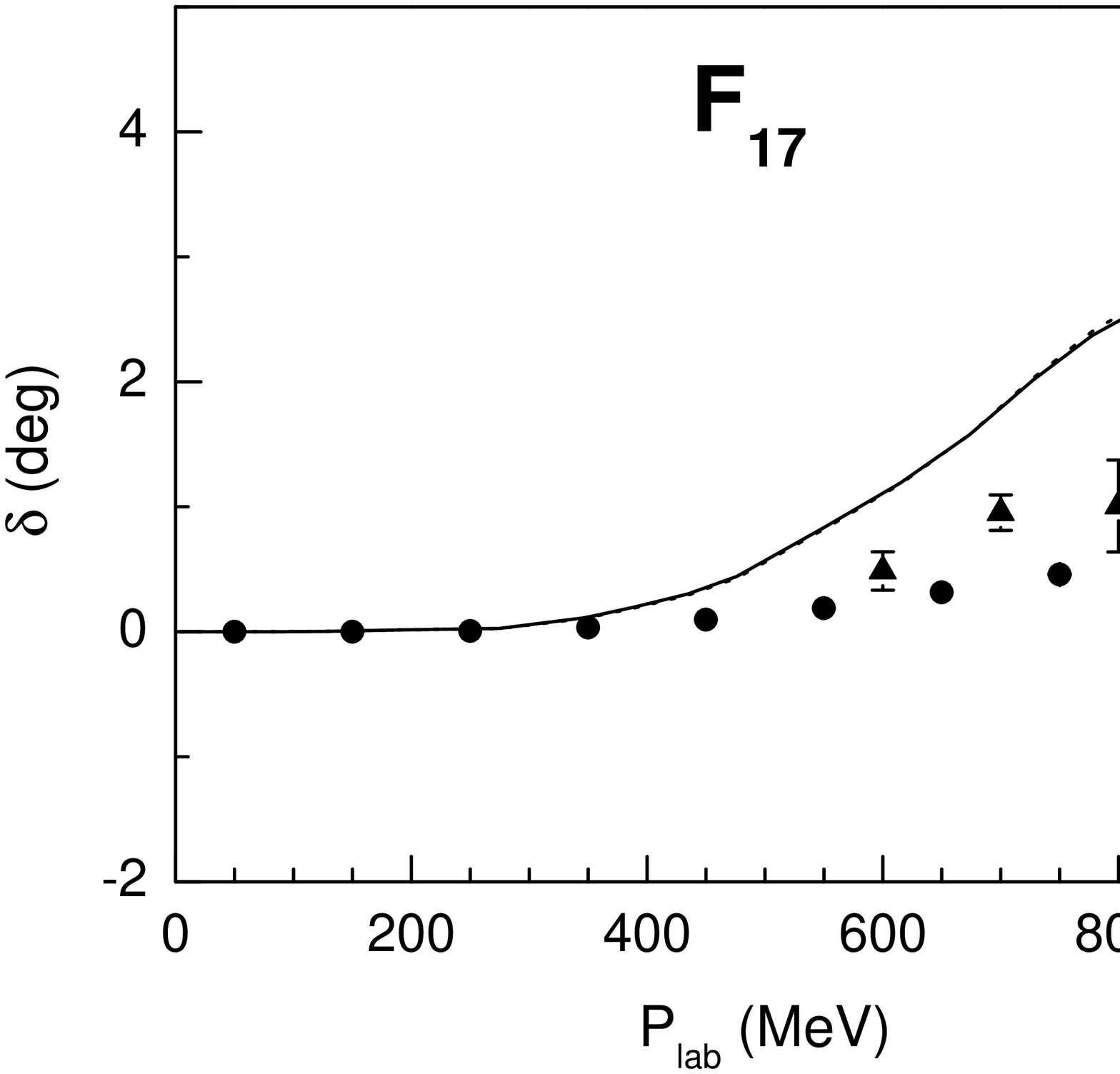,width=7.7cm} \vglue -2.3cm \caption{\small
\label{f0f1} $KN$ $F$-wave phase shifts. Same notation as in Fig.
\ref{s0s1}.}
\end{figure}

A RGM dynamical calculation of the $S$, $P$, $D$, $F$ wave $KN$
phase shifts with isospin $I=0$ and $I=1$ is made, and the
numerical results are shown in Figs. 1-4. Here we use the
conventional partial wave notation, the first subscript denotes
the isospin quantum number and the second one twice of the total
angular momentum of the $KN$ system. The solid lines represent the
results obtained by considering $\theta^S=35.264^\circ$ while the
dotted lines $-18^\circ$.

Actually, some authors have studied the $KN$ scattering processes
on a quark level. One recent work \cite{sle03}, a RGM calculation
including $\sigma$ and $\pi$ boson exchanges, gave an opposite
sign of the $S_{01}$ channel phase shifts. And another previous
work based on the constituent quark model concluded that a
consistent description of both isospin $S$-wave channels is not
possible \cite{bsi97}.  From Fig. \ref{s0s1} one can see that we
obtain the right sign of $S_{01}$ channel phase shifts, and our
results are in agreement with the experimental data for both
isospin $I=0$ and $I=1$ channels, although for $S_{11}$ they are a
little repulsive.

For higher angular momentum results (Figs. 2-4), comparing with
the recent RGM calculation of Lemaire {\it et al.} \cite{sle03},
we now get correct signs of $P_{11}$, $P_{03}$, $D_{13}$,
$D_{05}$, $F_{15}$, and $F_{07}$ waves, and for $P_{01}$,
$D_{03}$, and $D_{15}$ channels we also obtain a considerable
improvement on the theoretical phase shifts in the magnitude. We
also compare our results with those of the previous work of Black
\cite{nbl02}. Although our calculation achieves a considerable
improvement for all partial waves, the results of the $P_{13}$
channel are too repulsive in both Black's work and our present
one. The effects of the coupling to the inelastic channels and
hidden color channels are expected to be considered in future
work.

Since the annihilation interaction is not clear, its influence on
the phase shifts should be examined. We omitted the annihilation
part entirely to see its effect, and found that the numerical
phase shifts only have very small changes. This is because in the
$KN$ system the annihilations to gluons and vacuum are forbidden
and $u(d)\bar s$ can only annihilate to a kaon meson. This
annihilation part originating from $S$-channel acts in the very
short range, so that it plays a nonsignificant role in the $KN$
scattering process.

From the above discussion, one sees that our theoretical $S$, $P$,
$D$, $F$ wave $KN$ phase shifts are all reasonably consistent with
the experimental data. In this sense we can conclude that our
chiral SU(3) quark model can also be applied for the $KN$ system
in which an antiquark $\bar{s}$ is there besides four $u(d)$
quarks. Moreover, some information of the interactions between
quark-quark and quark-antiquark has been obtained which is useful
for developing this model to study the other hadron-hadron
systems.

\subsection{$\Delta K$ states}

In Ref. \cite{ssa04} a $\Delta K$ resonance state with $L=0$ and
$I=1$ has been obtained near the threshold. In the present work we
perform a RGM dynamical calculation to study the structures of
$\Delta K$ states in the framework of our chiral SU(3) quark
model.

\begin{figure}[htb]
\epsfig{file=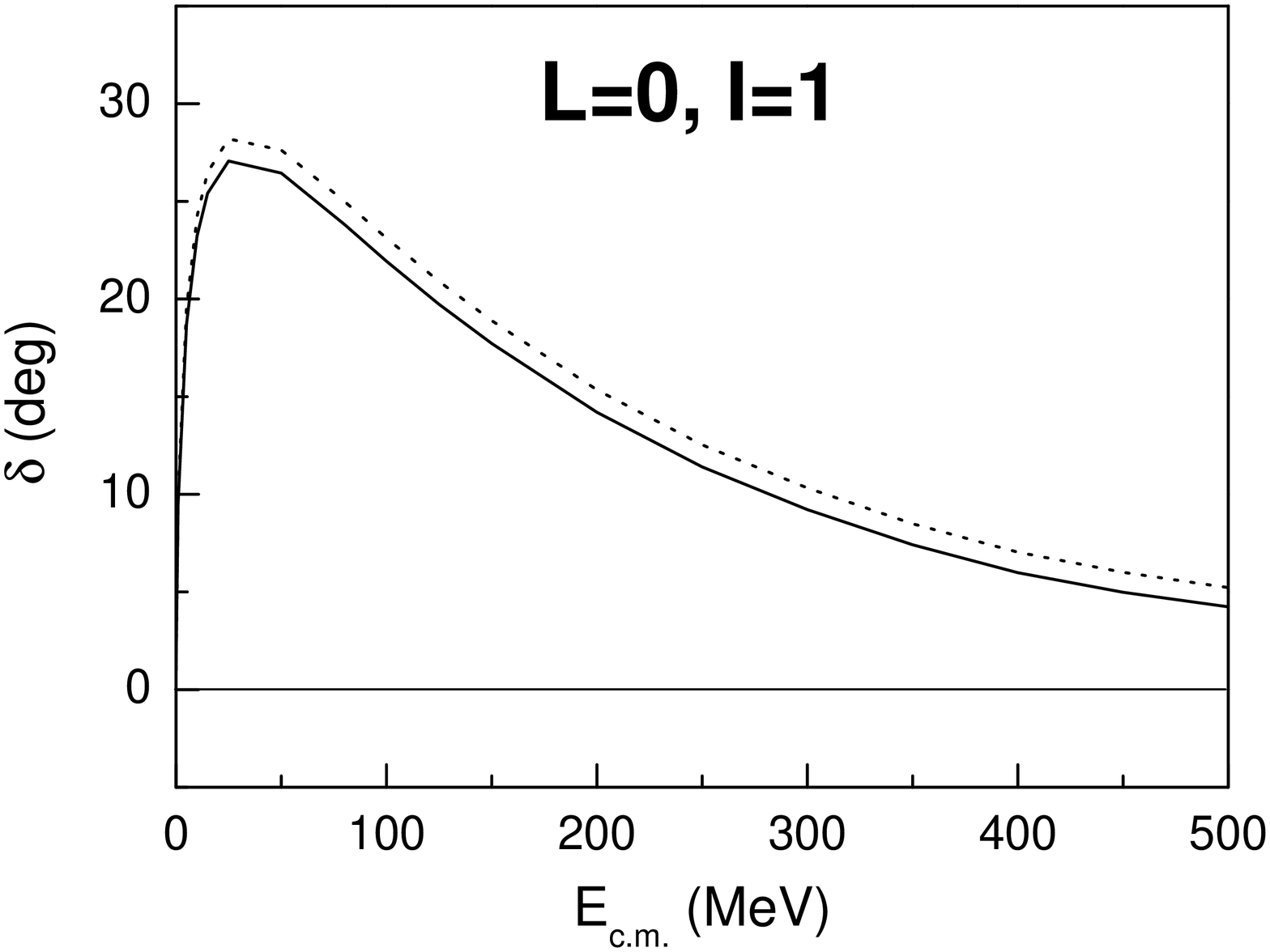,width=7.7cm}
\epsfig{file=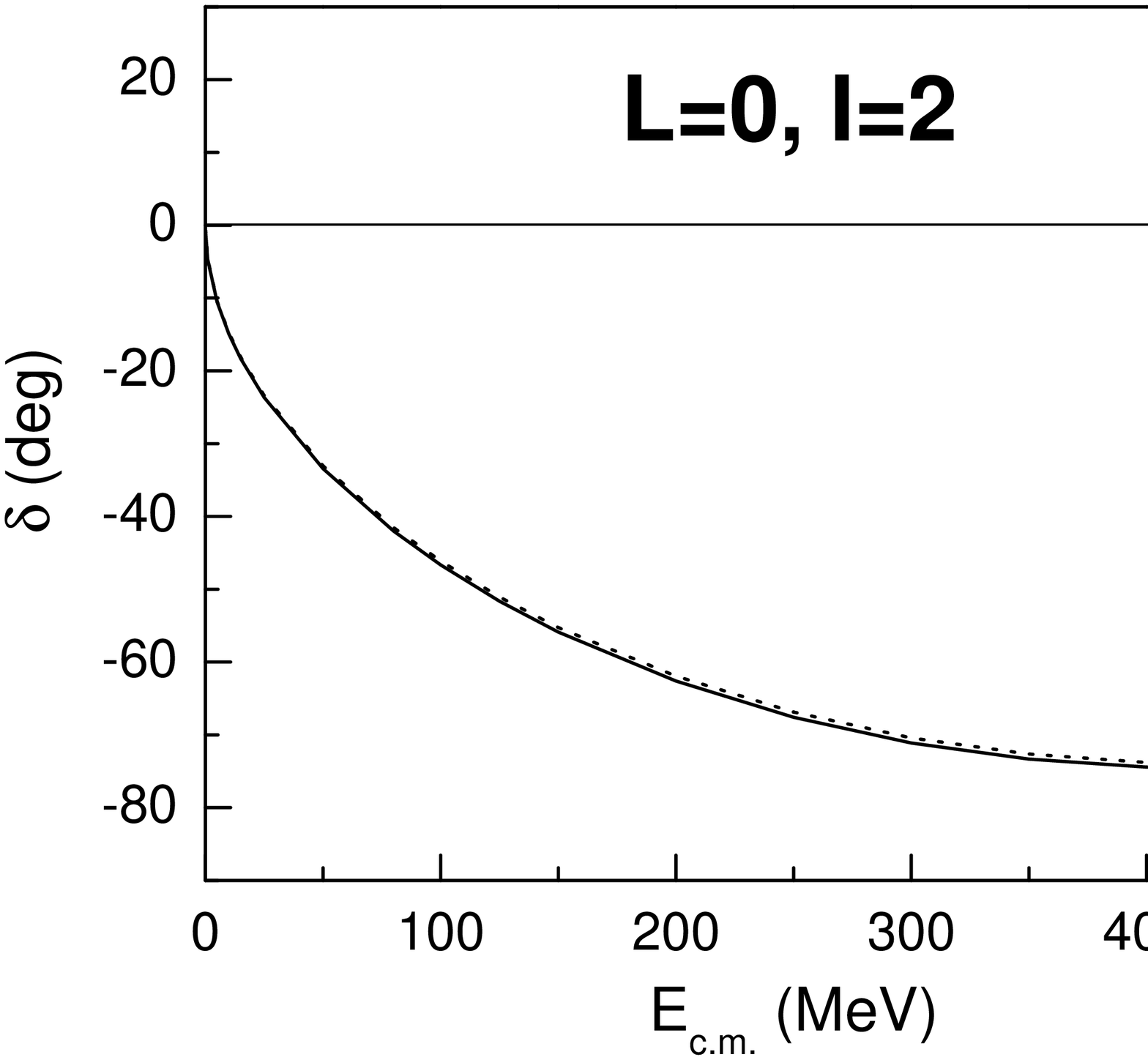,width=7.7cm} \vglue -2.3cm \caption{\small
\label{dks1s2} $\Delta K$ $S$-wave phase shifts as a function of
the energy of center of mass motion. The solid lines represent the
results obtained by considering $\theta^S=35.264^\circ$ while the
dotted lines $-18^\circ$.}
\end{figure}

First we calculated the $S$-wave $\Delta K$ phase shifts using two
sets of parameters fitted by the $KN$ phase shifts calculations,
although there is no experimental data to be compared. The results
are shown in Fig. \ref{dks1s2}. From which we can see that the
phase shifts are positive for isospin $I=1$ channel while negative
for $I=2$. This means the interaction between $\Delta$ and $K$ is
attractive in the $L=0$, $I=1$ state, and repulsive in the $L=0$,
$I=2$ state. Although in our calculations there is no obvious
$I=1$ $\Delta K$ resonance state, the $\Delta K$ $L=0$, $I=1$
state is really an interesting case, because it has attractive
interaction and when some other effects, such as coupling channel
effect, are considered, it might be a resonance or even a bound
state.

\begin{figure}[htb]
\epsfig{file=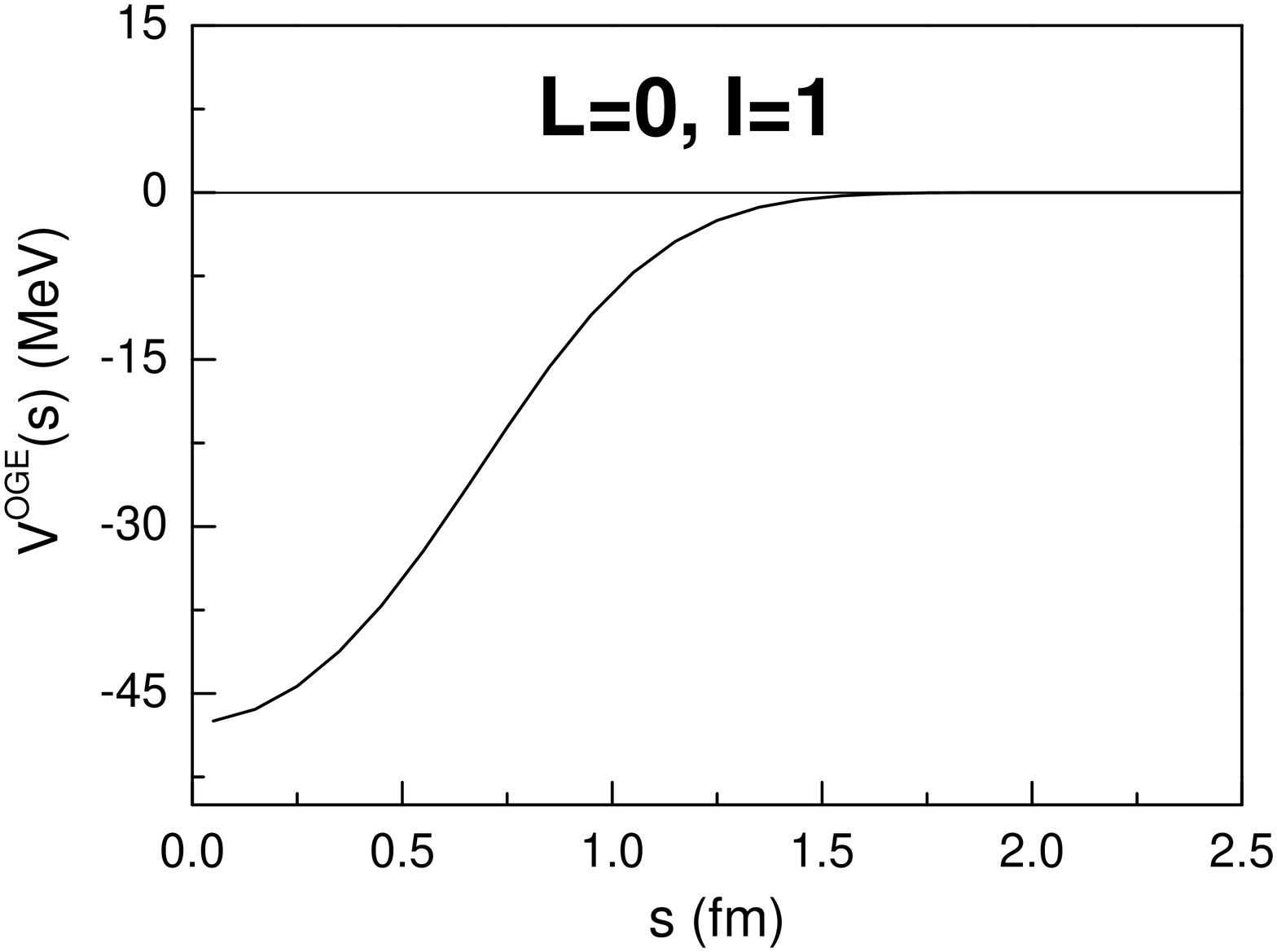,width=7.7cm}
\epsfig{file=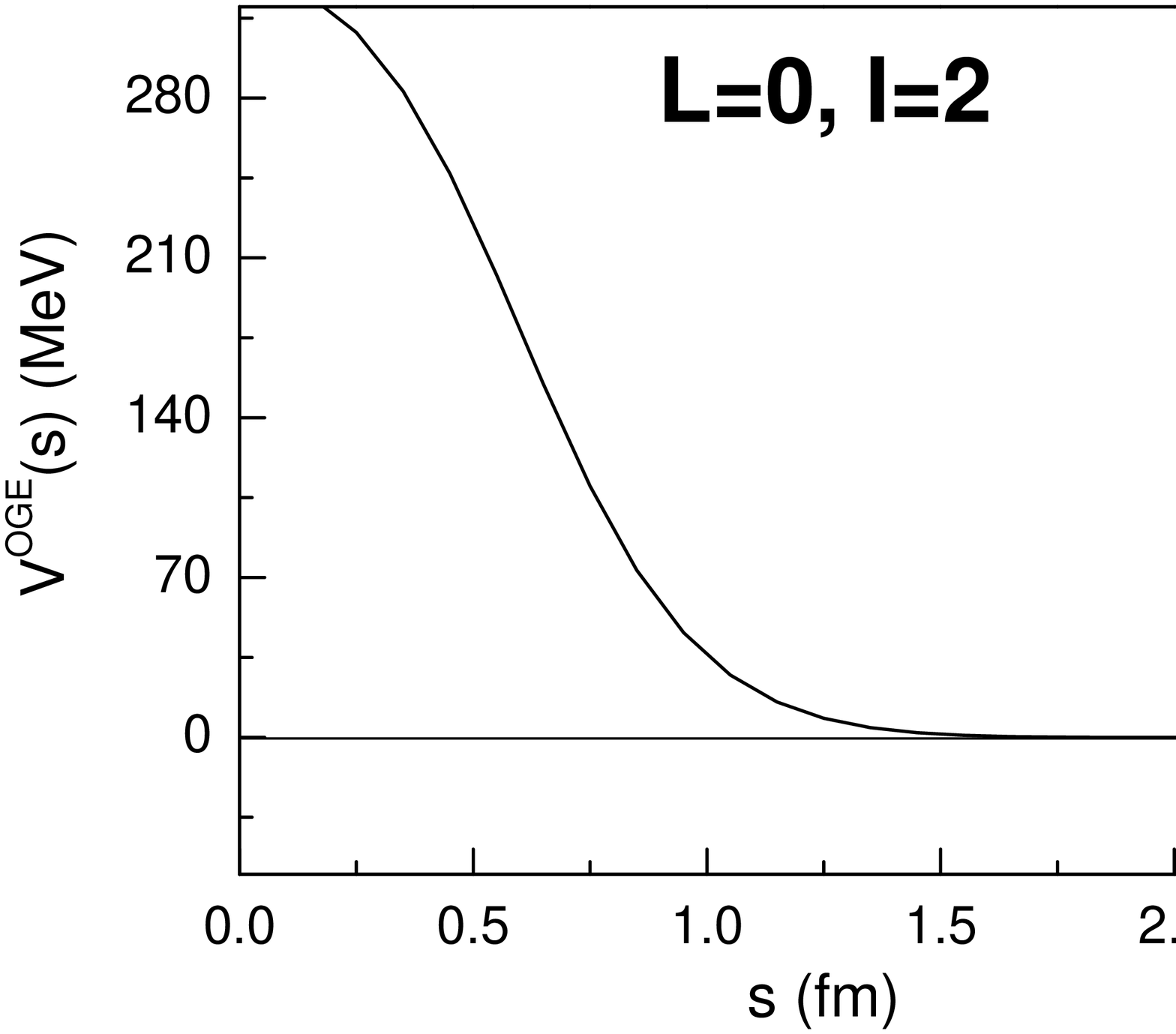,width=7.7cm}
\vglue -2.3cm
\caption{\small
\label{dkt1t2} The GCM matrix elements of OGE.}
\end{figure}

We did an analysis to see the contributions from different parts
of the interactions in the $\Delta K$ states. Fig. \ref{dkt1t2}
shows the diagonal matrix elements of the one-gluon-exchange
potential in the generator coordinate method (GCM) calculation
\cite{kwi77}, which can describe the interaction between two
clusters $\Delta$ and $K$ qualitatively. In Fig. \ref{dkt1t2}, $s$
denotes the generator coordinate and $V^{OGE}(s)$ is the OGE
effective potential between the two clusters. One sees that the
one-gluon-exchange potential is attractive for $I=1$ while
strongly repulsive for $I=2$. This means the OGE interaction plays
an important role in the $\Delta K$ system.

In our calculation, the tensor force is also included. Since the
kaon meson is spin zero, the tensor force of pion exchange, which
plays an important role in reproducing the binding energy of
deuteron \cite{lrd03}, now nearly vanishes and the nondiagonal
matrix elements between $S$ wave and $D$ wave offer pimping
contributions to the $\Delta K$ system.

To study the existence of a resonance or a bound state of the
$\Delta K$ system, we solve the RGM equation for the bound state
problem. The results show that if the parameters are taken to be
the values fitted by $KN$ phase shifts, the $\Delta K$ states are
unbound for both $I=1$ and $I=2$, though the $\Delta K$ state with
$L=0$ and $I=1$ has attractive interaction. However when we take
those parameters determined by the nucleon-nucleon phase shifts of
different partial waves and the hyperon-nucleon cross sections, we
get a weakly bound state of $\Delta K$ with about 2 MeV binding
energy for $I=1$ while unbound for $I=2$ in the present
one-channel calculation. Certainly the channel coupling between
($\Delta K$)$_{LSJ=0\frac{3}{2}\frac{3}{2}}$ and
($NK^*$)$_{LSJ=0\frac{3}{2}\frac{3}{2}}$ should be considered
further, because the effect of quark exchange between these two
channels is remarkable. It's expected that the energy of the
$\Delta K$ $L=0$ and $I=1$ state will be decreased once the
channel coupling is considered.

Here we would like to mention that our results of both $KN$ phase
shifts and $\Delta K$ states are independent of the confinement
potential in the present one-channel two-color-singlet-cluster
calculation. Thus our numerical results will almost remain
unchanged even the color quadratic confinement is replaced by the
color linear one.

\section{Conclusions}

In this paper, the chiral SU(3) quark model has been extended to
the system with an antiquark, and the $S$, $P$, $D$, $F$ wave $KN$
phase shifts have been studied by solving the resonating group
method (RGM) equation based on this model. Comparing with another
RGM calculation \cite{sle03}, we can obtain correct signs of
$S_{01}$, $P_{11}$, $P_{03}$, $D_{13}$, $D_{05}$, $F_{15}$ and
$F_{07}$ wave phase shifts, and a considerable improvement on the
theoretical phase shifts in the magnitude for $P_{01}$, $D_{03}$
and $D_{15}$ channels. It turns out that our chiral SU(3) quark
model is quite successful to be extended to study the $KN$ system,
in which an antiquark $\bar{s}$ is there besides four $u(d)$
quarks. At the same time some useful information of quark-quark
and quark-antiquark interactions is provided.

Also, we have studied the $\Delta K$ systems using our model. If
the parameters are taken to be the values fitted by $KN$ phase
shifts, the $S$-wave $\Delta K$ states for both $I=1$ and $I=2$
are unbound, though the interaction between $\Delta$ and $K$ is
attraction for the $I=1$ state. However when we take the
parameters determined by the $NN$ phase shifts and the $YN$ cross
sections, we find there is a weakly $\Delta K$ bound state with
$L=0$ and $I=1$, and the binding energy is about 2 MeV in our
present one-channel calculation.

To examine if ($\Delta K$)$_{LSJ=0\frac{3}{2}\frac{3}{2}}$ is
possible to be a resonance or a bound state, the channel coupling
between ($\Delta K$)$_{LSJ=0\frac{3}{2}\frac{3}{2}}$ and
($NK^*$)$_{LSJ=0\frac{3}{2}\frac{3}{2}}$ would be considered in
our future work.

\begin{acknowledgements}
One of the authors (F. Huang) is indebted to Prof. Y.B. Dong and
Dr. D. Zhang for a careful reading of the manuscript. This work
was supported in part by the National Natural Science Foundation
of China No. 90103020.
\end{acknowledgements}

\end{document}